\documentclass[12pt]{iopart}
\usepackage{graphicx}
\begin{document}
\title{Balanced superconductor-insulator-superconductor mixer on a 
9~$\mu$m silicon membrane}
\author{M P Westig, K Jacobs, J Stutzki, M Schultz, M Justen and 
C E Honingh}
\address{KOSMA, I.~Physikalisches Institut, Universit\"at zu K\"oln, 
50937 K\"oln, Germany}
\ead{westig@ph1.uni-koeln.de}
\begin{abstract}
We present a $380-520$~GHz balanced 
superconductor-insulator-superconductor (SIS) mixer on 
a single silicon substrate.
All radio-frequency (RF) circuit components are fabricated on a 
$9~\mu$m~thick membrane. The intermediate frequency (IF) 
is separately amplified and combined.
The balanced mixer chip, using Nb/Al/Al$_{2}$O$_{3}$/Nb SIS junctions, 
is mounted in a tellurium copper waveguide block at 4.2~K using 
Au beam lead contacts. We find uncorrected minimum receiver 
double-sideband noise temperatures of 70~K and a 
noise suppression of up to 18~dB, measured within a $440-495$~GHz 
RF and a $4-8$~GHz IF bandwidth, representing state-of-the-art device
performance.
\end{abstract}
\submitto{\SUST}
\maketitle
\section{Introduction}
Astronomical discoveries as well as better insight into
already known astronomical phenomena are often 
accompanied by new or improved detector technologies 
\cite{blain2002}. 

In the submillimetre (submm) regime the implementation of 
superconductor-insulator-superconductor (SIS) heterodyne 
mixers as replacement for earlier semiconductor (Schottky) 
mixers has resulted in receivers with a sensitivity of a 
few times the quantum limit. Further improvement of observation 
efficiency is achieved by using SIS mixers in focal plane arrays.

Balanced superconductor-insulator-superconductor (SIS)
heterodyne frequency mixers are advanced devices 
for receivers in the submm regime. They provide a separate
local oscillator (LO) port, useful for building arrays, do not
limit the intermediate frequency (IF) bandwidth by optical
diplexing and suppress LO amplitude (AM)
noise and thermal noise incident at 
the LO port \cite{maas1986}.

Chattopadhyay~\emph{et al}~\cite{chattopadhyay1999} 
realized an integrated quasioptical balanced mixer with
a double-sideband (DSB) receiver noise temperature 
$T_{rec,DSB}=$105~K at 528~GHz and 
an IF bandwidth of 0.5~GHz in 1999. In 2000, 
Kerr \emph{et al}~\cite{kerr2000} reported an integrated 
waveguide-coupled $200-300$~GHz balanced mixer on a 
quartz substrate with an IF bandwidth of $1-2$~GHz.
Between $225-300$~GHz the authors measured the 
receiver noise temperature and the noise suppression.
They obtained $T_{rec,DSB}=$ $46-78$~K and the 
noise suppression was better than 12~dB over the measurement
bandwidth. In 2008, Serizawa \emph{et al} \cite{serizawa2008} 
published results of a $400-495$~GHz waveguide branch-line 
balanced mixer with $T_{rec,DSB}=55-120$~K and an IF 
bandwidth of $4-8$~GHz. At the Caltech Submillimetre 
Observatory (CSO), four waveguide branch-line 
balanced mixers covering the $180-720$~GHz atmospheric
windows with $4-8$~GHz IF bandwidth are under development
\cite{kooi2004}.
The Group for Advanced Receiver Development 
(GARD, Chalmers University of Technology) 
demonstrated a balanced waveguide hot-electron bolometer (HEB)
mixer covering the RF band $1.25-1.39$~THz and with 
an IF bandwidth of $3$~GHz. Over the RF band the averaged 
DSB receiver noise temperature was better than 
1200~K \cite{meledin2009}.

Receiver development for e.g.~the 
Atacama Large Millimetre Array (ALMA) \cite{wootten2009} 
stimulated the development of single-pixel 
superconducting side-band separating heterodyne 
mixers where waveguide circuit elements, 
like branch-line couplers, are implemented \cite{ALMA}. 
The technological complexity of fabricating 
waveguide branch-line couplers increases with frequency.
E.g.~typical spacing distances of the 
coupler branches in \cite{meledin2009} do not exceed 
$35~\mu$m. Therefore, instead of using e.g.~milling techniques, 
photolithography together with fine copper electroplating 
was used to fabricate this branch-line coupler.
Next generation instruments where advanced mixer 
devices, such as balanced or single-sideband mixers, 
will be used in focal plane arrays would greatly benefit 
from a frequency scalable technology where key 
elements of these circuits (e.g.~branch-line couplers 
or LO distribution circuits) are integrated on one 
chip. This would provide the possibility to produce 
focal plane array receivers (i.e.~many pixels) 
in large numbers and with high yields. 

In this paper we present a balanced SIS mixer on a silicon (Si) 
membrane. Si membrane technology and 
microfabrication techniques provide a possibility
to overcome the difficulties of realizing balanced SIS
mixers at 1 THz and above, provided that a suitable
superconductor detector and transmission line technology 
is used. $T_{rec,DSB}$ and noise rejection have been measured 
within the $440-495$~GHz bandwidth of our LO 
over an IF bandwidth of $4-8$~GHz. 
The influence of thermal radiation superimposed 
with the LO signal on the mixer IF output power 
is measured, illustrating
the mechanism of AM and thermal noise rejection.
We show that $T_{rec,DSB}$ 
does not change when DC-biasing the balanced circuit 
inversely or with the same polarity if the corresponding
$\Sigma$ or $\Delta$ port of the external $180^{\circ}$ 
IF hybrid is used. Building on the present paper we wish
to present a detailed analysis in another publication. 
The expected performance of the mixer
chip can be determined from the quantum theory of 
mixing \cite{tucker1979,tucker1985} and results of 
our electromagnetic design. 
By comparing this analysis to the measurements
we expect to determine the influence of e.g.~fabrication 
tolerances and to identify improvements that 
can be implemented in future design iterations.  
\section{Working principle and design of the balanced SIS mixer}
\begin{figure}[t]
\includegraphics[width=\textwidth]{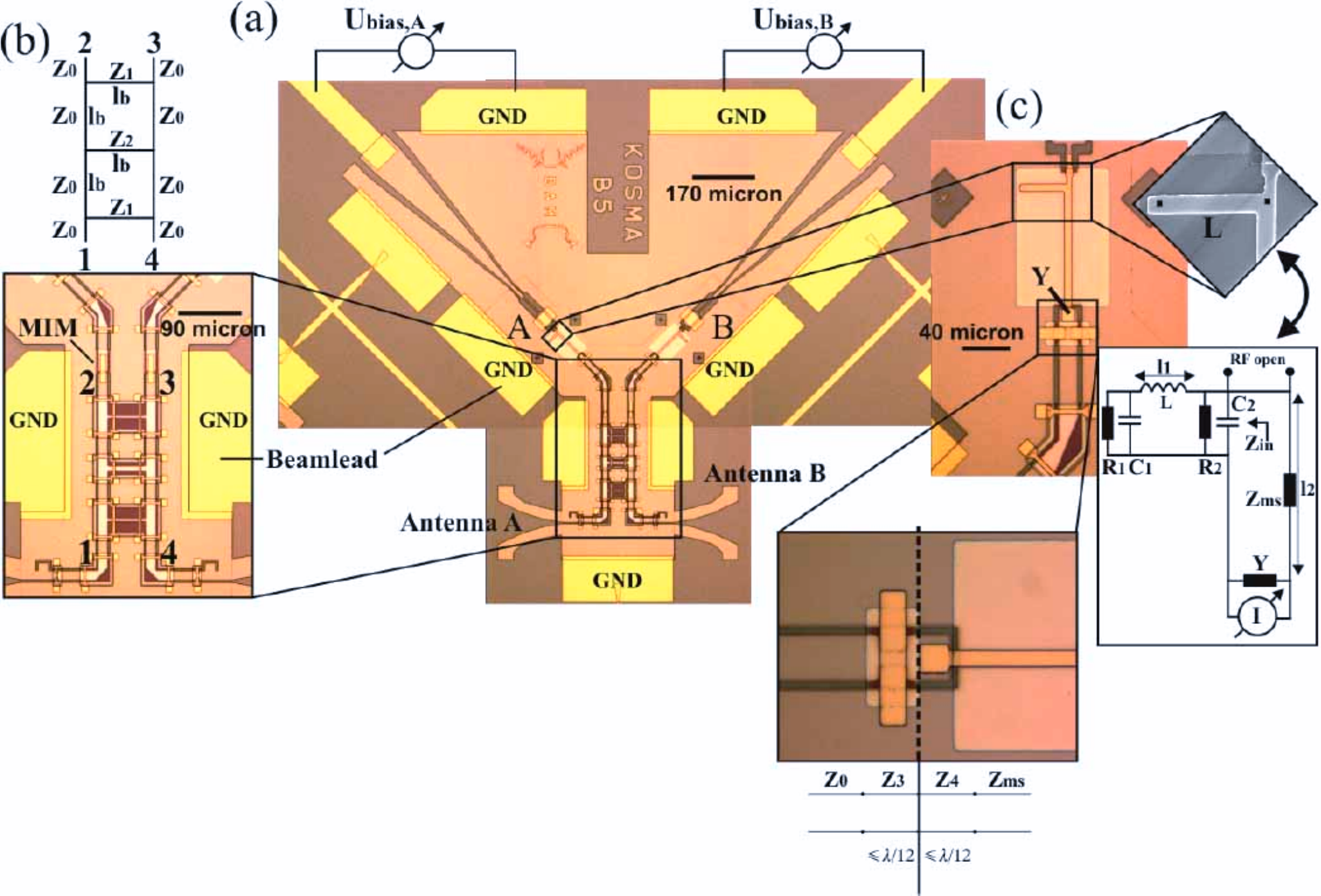}
\caption{\label{fig01}(Colour online) (a) shows a balanced 
mixer device on the handle wafer before substrate
geometry definition. Two tapered slotline waveguide antennas are
connected to the 90$^\circ$~CPW 
branch-line coupler, magnified in (b) feeding
two pairs of SIS junctions with one pair magnified in (c). Each
twin-junction pair is connected to a separate RF blocking filter. 
Beam lead contacts provide ground and signal connections and hold 
the device suspended inside the mixer block [figure~\ref{fig02}(a)]. 
In (c) the alternated line transformer
is indicated by a box around the region where we specify the circuit's
admittance $Y$. The section with impedance
$Z_3$ is realized with an inverted microstrip transmission 
line whereas the sections with impedances $Z_0$ and $Z_4$ 
are CPW transmission lines. The microstrip line 
with impedance $Z_{ms}$ is the input section of 
the quarter-wave transformer which is part of
the SIS junction tuning circuit.}
\end{figure}

\begin{figure}[t]
\includegraphics[width=\textwidth]{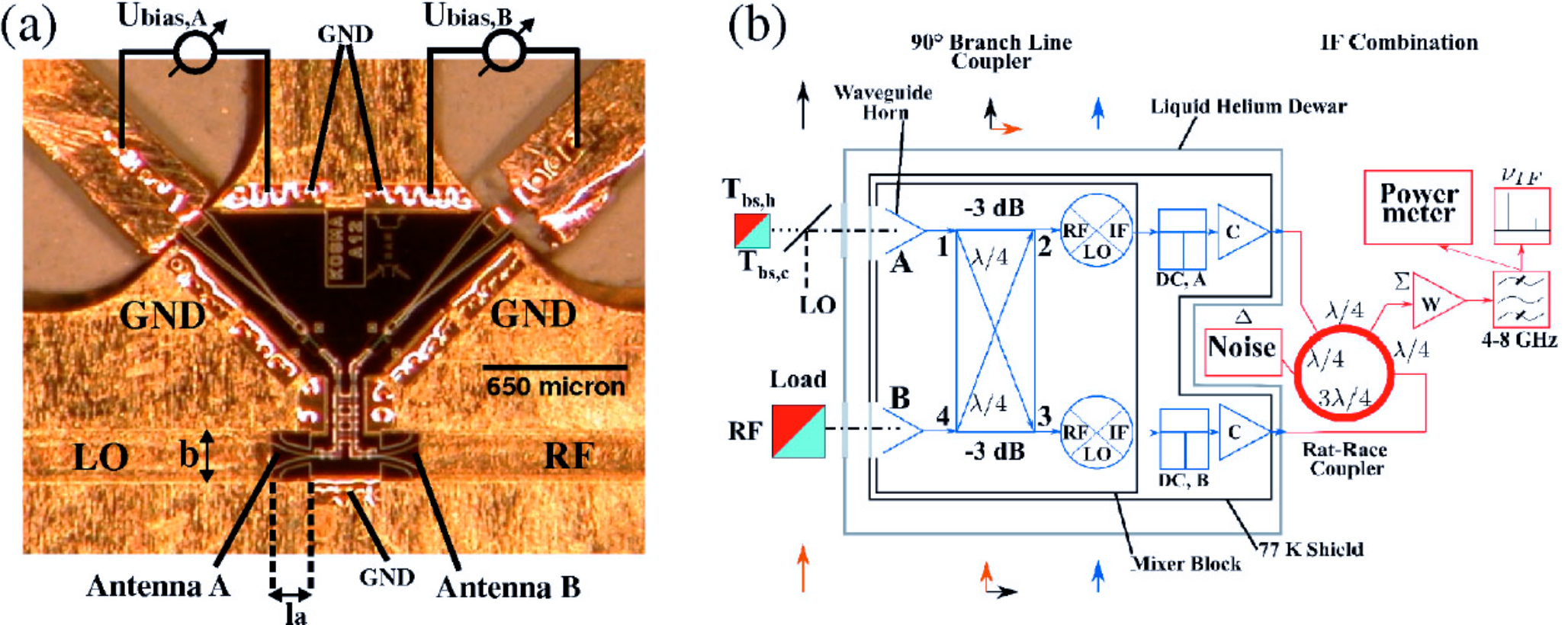}
\caption{\label{fig02}(Colour online)
(a) shows a balanced mixer device mounted inside the waveguide
mixer block.
(b) shows a scheme of the experimental setup and the 
signal path diagram of the balanced
mixer. Blue parts are at 4.2~K while red parts are at 295~K.  
The rectangles represent calibration blackbodies
$T_{h}$ (295~K, red) and $T_{c}$ (77~K, light blue). 
Arrows indicate the phase of RF, LO and IF, 
respectively.
}
\end{figure}

We start with a brief discussion of the working principle of the balanced
SIS mixer device shown in figure \ref{fig01}(a).
The circuit is realized using superconducting Nb transmission lines.
Two tapered slotline antennas \cite{lin2001} A and B
are matched via a slotline shorted stub to a $Z_0=43~\Omega$
coplanar waveguide (CPW) each connected to the input of a
90$^\circ$~CPW branch-line coupler with the same input 
impedance [magnified in figure \ref{fig01}(b)].
At CPW discontinuities,
the odd mode of the propagating TEM wave is suppressed 
by $3~\mu$m wide ground connecting Nb bridges on top of a
900~nm $\mathrm{SiO_2}$ layer that connect the ground 
planes at either side of the CPW.
Numbers $1-4$ label the coupler ports in figure \ref{fig01}(b) and 
figure \ref{fig02}(b).
Signals received by antenna A (guided to 
port 1) and antenna B (guided to port 4) are 
equally distributed to port 2 and 3
with a relative phase difference of $-\pi/2$ and $+\pi/2$,
respectively.
This is achieved by three branches separated 
by $l_b=\lambda/4$. All branches have a length of
$l_b$ with impedances $Z_1=Z_0\cdot 2.415$
and $Z_2=Z_0\cdot\sqrt{2}$, cf.~\cite{reed1956}.
The two output ports of the coupler with impedance $Z_0$ 
are each connected to a SIS microstrip tuning circuit 
via an uniplanar alternated line \cite{bramham1961} 
CPW-to-microstrip transformer. 
The uniplanar alternated line transformer section of the balanced circuit 
together with a circuit schematic is shown in figure \ref{fig01}(c). 
The impedance $Z_0$ of the
output ports of the branch-line coupler is matched to the
input impedance $Z_{ms} = 21~\Omega$ of the microstrip 
line. For $Z_0 = Z_4$ and $Z_{ms} = Z_3$ 
(i.e.~with alternating impedances) the electrical length of the 
two transformer sections become 
\begin{equation}
\frac{1}{2\pi}\cdot\mathrm{arctan}\left[
\frac{Z_0/Z_{ms}}{(Z_0/Z_{ms})^2 + Z_0/Z_{ms} + 1}\right].
\label{eq01}
\end{equation}
For $Z_0 = Z_{ms}$, (\ref{eq01}) equals exactly $1/12$ 
and for $Z_0 \not= Z_{ms}$ it takes a smaller value. 

Figure \ref{fig01}(c) shows one of the tuning circuits and 
a SIS junction pair (black dots in the inset indicate
a junction). The circuit schematic explains the superconducting 
tuning circuit. A microstrip inductance 
$L$ with length $l_1$ compensates $C_1$ and $C_2$ resulting in 
$Z_{in}=R_1R_2\cdot(R_1+R_2)^{-1}$, $R_1$ and 
$R_2$ being the normal state resistance of the two SIS junctions. Here
$Z_{in}=10~\Omega$ and $R_1=R_2$. A microstrip line with length
$l_2=\lambda/4$ and input impedance $Z_{ms}$ matches 
$Z_{in}$ to the admittance $Y$ of the RF structure. 
Finally, in the SIS junctions 
RF and LO signal are multiplied to produce 
IF signal $\nu_{IF}=\left|\nu_{RF} -\nu_{LO}\right|$. 
A CPW RF choke (filter) is used to block further 
propagation of the RF and LO signals 
and provides a bandpass for IF signals. 

Figure \ref{fig02}(a) 
shows device $\mathcal{A}$12, used for our measurements, 
assembled inside the mixer block. Antenna A and B
with length $l_a=170~\mu$m extend into a split 
block full-height waveguide cut in the $\vec{E}$-plane.
The waveguide dimensions are $b=230~\mu$m and $a=2b$ ($a$ is 
normal to the paper plane) with a step of $131~\mu$m
at the antenna feedpoint in $a$-direction and in both 
mixer block halves.
The top face as well as the bottom face of the device is 
separated from the mixer block wall
by a gap of $90~\mu$m [figure \ref{fig02}(a)]. 
The IF signal output beam leads are connected to a 
circuit board transmitting the IF and allowing 
each junction pair to be biased separately. 
A magnetic field, produced by an electromagnet which 
is attached to the mixer block, is 
used to suppress Cooper pair tunneling.
Two corrugated horn antennas [not shown in figure \ref{fig02}(a)] 
are attached at both waveguide inputs 
of the mixer block to couple the free space LO and RF
signal to the waveguides.
\section{Device fabrication}
The device is fabricated in-house 
on Silicon-On-Insulator (SOI) wafers with a 
high-resistivity ($>$10~k$\Omega$-cm) device layer of 9~$\mu$m
thickness (cf.~also \cite{kaul2004}).
The Nb/Al/$\mathrm{Al_2O_3}$/Nb 
tunnel layers are deposited with DC-magnetron sputtering
and patterned by UV optical lithography.
The 1~$\mu \mathrm{m}^2$ area junction top electrodes are
defined by electron-beam lithography. 
The 300 nm sputtered $\mathrm{SiO_2}$ 
dielectric layer is defined with standard self-aligned liftoff.
A second $\mathrm{SiO_2}$ layer of 600~nm is added in the area of the
Nb ground connection bridges loading 
the CPW. The final Nb wiring layer is
400 nm thick and is again defined by UV optical lithography.
After defining the beam lead 
contacts with a sputtered gold seed layer, 
the beam leads 
are electrolytically gold plated to a thickness of 2.5~$\mu$m. The handle wafer 
bulk silicon is removed from the backside in an inductively coupled
plasma deep reactive ion etch (DRIE) step stopping on 
the SOI buried oxide (BOX) layer.
Subsequently the device substrate geometry 
is defined on backside by photoresist lithography, 
and the individual devices are etched out with RIE 
to remove the BOX layer followed by an anisotropic 
Bosch Si etch recipe for the silicon device layer.
\section{Experimental setup}
The mixer block is mounted in a cold optics assembly 
on a 4.2~K liquid helium dewar stage. 
The LO signal passes through a $443~\mu$m thick 
teflon window and the separate RF window is made of
a plane slab of $505~\mu$m HDPE. No special infrared coating
or grooving was used on this window which has a measured
transmission of $0.95$ at $440$~GHz decreasing to
$0.87$ at $495$~GHz.  
Infrared radiation is blocked by
$200~\mu$m thin HDPE windows on the dewar 77~K shield.
Radiation emitted by blackbody sources at temperature 
$T_{h}=295$~K or $T_{c}=77$~K (load) 
is received by antenna B [figure \ref{fig02}(b)]. 
The LO signal is coupled to the device
by a $25~\mu$m thin Mylar beamsplitter at 295~K. Behind
the beamsplitter, thermal radiation from a load at temperature 
$T_{bs,h}=T_{h}$ or $T_{bs,c}=T_{c}$
is superimposed with the LO signal and 
received by antenna A [figure \ref{fig02}(b)]. 
Two cryogenic low-noise
MMIC WBA13 amplifiers \cite{wadefalk2005} 
[$T_{amp}\approx 4$~K over $4-12$~GHz, $C$ in 
figure \ref{fig02}(b)] and two $4-10$~GHz bias-T circuits
are connected to the balanced mixer IF ports through the
circuit board shown in figure \ref{fig02}(a).
Both amplifiers provide an equal gain 
throughout all measurements in this paper.
A low noise warm amplifier [W in figure \ref{fig02}(b)] 
is used for further amplification.
The combined IF output power at the output of
a $180^{\circ}$ IF hybrid \cite{ETindustries} 
[we use a rat-race coupler outside of the liquid helium 
dewar, figure \ref{fig02}(b)] is
averaged over $4-8$~GHz and measured with a
power meter or is measured as a 
function of IF frequency with a spectrum analyzer.
Thanks to the anti-symmetrical IV curve
of a SIS junction, an additional phase shift of 
$\pi$ can be added to the respective IF signals by 
using an inverse polarity DC bias for the two SIS junction pairs. 
This enables the use of the $\Sigma$ port of the 
$180^{\circ}$ IF hybrid that combines the two IF signals
for the balanced IF output.
Noise added to the LO signal
cancels at the $\Sigma$ port and is measured at the
$\Delta$ port [figure \ref{fig02}(b)]. This situation is reversed 
by operating the two SIS junction pairs with equal bias 
polarity where now $\Sigma$ is the
noise port and IF signals are added at $\Delta$.
\section{Noise temperature and noise rejection measurement}
\begin{figure}[t]
\includegraphics[width=1\columnwidth]{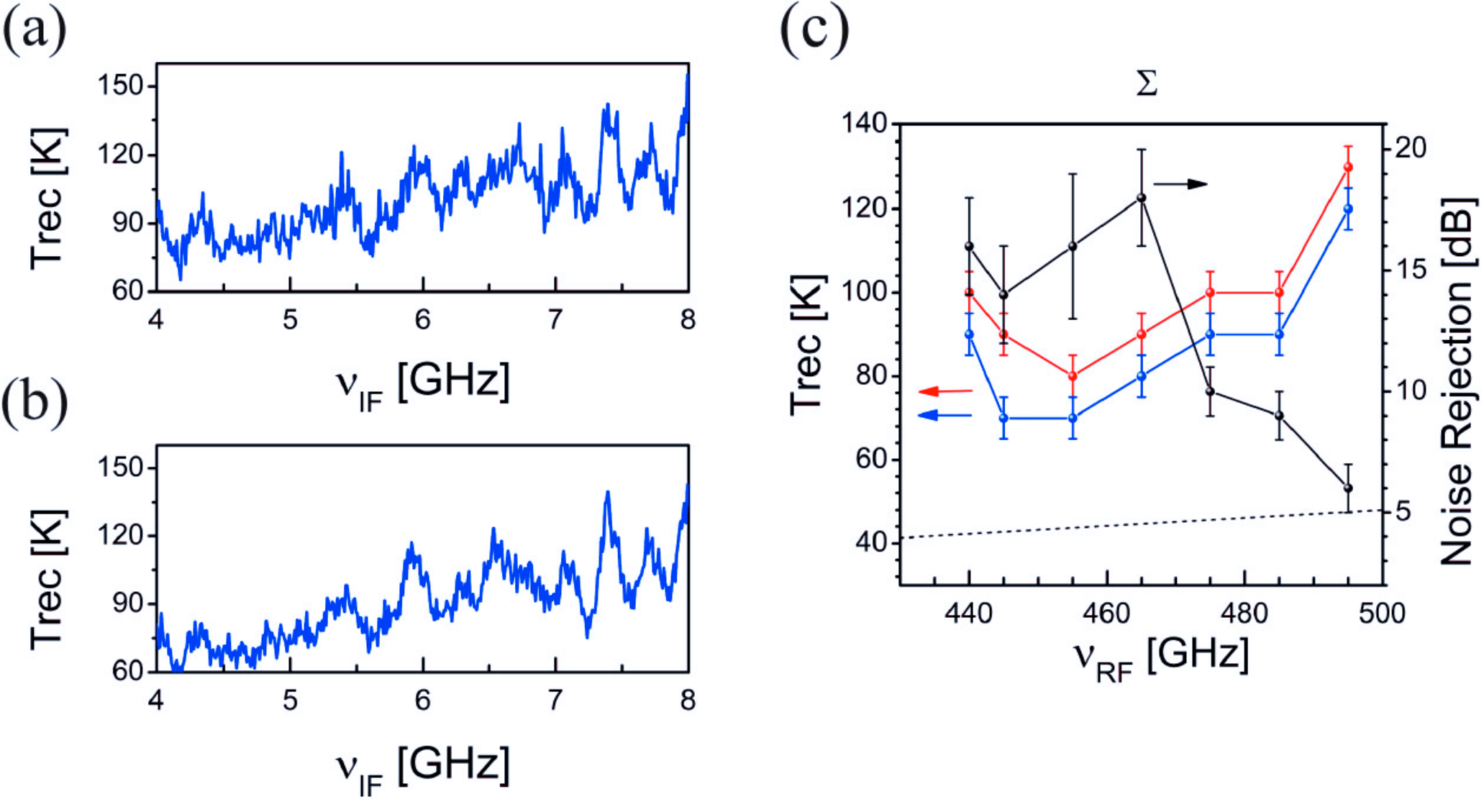}
\caption{\label{fig03}(Colour online) 
Left: $T_{rec,DSB}$ is measured as a function of IF 
for $U_{bias,A}=-U_{bias,B}=-1.55$~mV (a) and
$U_{bias,A}=U_{bias,B}=1.55$~mV (b). Trace (a) [(b)] is taken with
a spectrum analyzer at the $\Sigma$ [$\Delta$] 
port and with constant LO beamsplitter termination 
temperature $T_{bs,c}$ [figure \ref{fig02}(b)]. 
Right: $T_{rec,DSB}$ is measured with IF output power 
averaged over $4-8$~GHz as function of $\nu_{RF}$. The LO
beamsplitter termination temperature is at constant temperature 
$T_{bs,h}$ (red) or $T_{bs,c}$ (blue). The solid black curve is the mixer's noise
rejection and the dashed black line indicates $2h\nu_{RF}/k_B$, with
$h$ and $k_B$ the Planck and Boltzmann constant.}
\end{figure}
\begin{figure}[t]
\includegraphics[width=1\columnwidth]{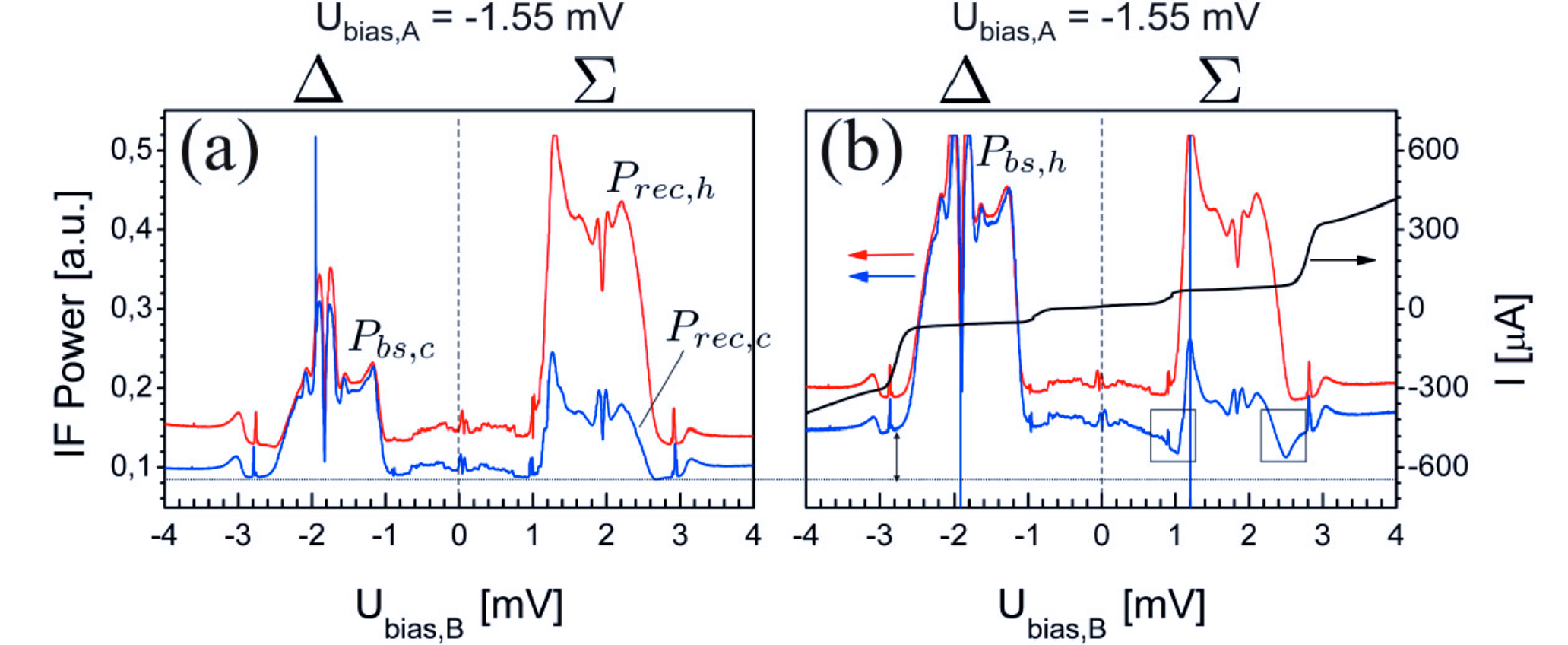}
\caption{\label{fig04}(Colour online) 
IF output power curves as a function of $U_{bias,B}$
($\nu_{RF}=455$~GHz) and the LO pumped SIS IV characteristic
of the junction pair B (black solid curve).
The IF output power is measured at the $\Sigma$ 
port as function of $U_{bias,B}$ [figure \ref{fig02}(b)]. 
In (a) the LO beamsplitter termination temperature is at 
constant temperature $T_{bs,c}$ and in (b) at $T_{bs,h}$. The
red (blue) curve was taken with a 295~K (77~K) load in front of the
RF port. For $U_{bias,B}<0$ the $\Sigma$ port acts like the $\Delta$ port
and the noise added to the mixer is measured while for
$U_{bias,B}>0$, $T_{rec,DSB}$ is measured at the $\Sigma$ port.
The dashed vertical line separates the $\Delta$ from the $\Sigma$
branch of the IF output power trace. In (b) the IF output power 
trace is shifted up with respect to the trace shown in (a) 
due to the excess of thermal noise incident at the LO port.}
\end{figure}
We determine $T_{rec,DSB}$ by measuring
the ratio $Y=P_{rec,h}/P_{rec,c}$, $P_{rec,h}$ and
$P_{rec,c}$ being the IF output power averaged
over the resolution bandwidth while a load temperature
$T_{h}$ and $T_{c}$ is placed in front of the
RF window [figure \ref{fig02}(b)]. The uncorrected 
DSB receiver noise temperature is 
$T_{rec,DSB}=\left(T_{h}-YT_{c}\right)/\left(Y-1\right)$, where 
$T_{h}$ and $T_{c}$ are the physical temperatures of the loads
(Rayleigh-Jeans limit). 
At 455~GHz we measured $T_{rec,DSB}$ 
as function of $\nu_{IF}$ (resolution: 3~MHz) 
with one junction pair biased inversely with
respect to the other pair [figure \ref{fig03}(a)] and with
positive bias polarity for both junction pairs 
[figure \ref{fig03}(b)].
In the latter case we measured 
no significant difference in $T_{rec,DSB}$. 
In figure \ref{fig03}(c) we show $T_{rec,DSB}$ at various LO frequencies.
The IF output power was averaged over the $4-8$~GHz IF output bandwidth.

In figure \ref{fig04}(a) and (b) 
IF output power is measured as function of
$U_{bias,B}$ with constant LO beamsplitter termination 
temperature $T_{bs,c}$ and $T_{bs,h}$ and
with constant voltage bias $U_{bias,A}=-1.55$~mV in
both measurements.
We denote with $U^{+,-}_{ph}$ voltages within the
range of the first photon step below the gap 
of the LO pumped SIS IV characteristic on the 
positive $(+)$ and the negative $(-)$ 
branch [figure \ref{fig04}(b)]. The thermal 
power $P_{bs,c}=k_{B}T_{bs,c}B$ and
$P_{bs,h}=k_{B}T_{bs,h}B$, superimposed with LO radiation and
acting as noise incident on the device, is measured in the
$\Delta$ branch of the IF output power trace for
$U_{bias,B}=U^{-}_{ph}$. Here, $B$ is the IF bandwidth of $4-8$~GHz.
Evidently for $T_{bs,c}$ less noise is added to the mixer
visible in the height of the IF power trace in the $\Delta$
branch.
In the $\Sigma$ branch the rejection of the 
noise is clearly visible, especially for
LO beamsplitter termination temperature 
$T_{bs,h}$. Here it is observed that the IF output power
trace is shifted downwards due to the noise rejection as
indicated in figure \ref{fig04}(b) by the region between the
rectangles. 
This occurs for $U_{bias,B}=U^{+}_{ph}$ and in this branch
we measure $T_{rec,DSB}$.
SIS junction pair A provides a constant IF output power 
during each 
voltage sweep resulting in a
relative shift between
the blue and the red trace in figure \ref{fig04}(a) and (b)
and in an overall shift of both traces in 
figure \ref{fig04}(b) (black arrow)
relative to the traces in figure \ref{fig04}(a).

The noise rejection of the balanced mixer is measured
by the ratio $dP^{+}/dP^{-}$, $dP^{+,-}$ being the difference
in IF output power between the red and blue line 
for $U_{bias,B} = +1.55$~mV $(+)$ and $U_{bias,B} = -1.55$~mV $(-)$ 
[figures \ref{fig04}(a), (b)]. 
The black solid line in figure \ref{fig03}(c) shows the noise rejection
in dB according to the formula 
$10~\mathrm{Log}\left(dP^{+}/dP^{-}\right)$.
We measured several times the noise rejection 
both for constant LO beamsplitter termination 
temperatures $T_{bs,h}$ and $T_{bs,c}$
and averaged the results. Error bars are the mean square deviation.

A conservative estimate of the contribution of the window 
and infrared filter to the measured noise temperature is 
$25-30$~K. 
Based on an average balanced mixer gain of $-4$~dB in the best 
part of the RF band, the IF noise contribution is $10$~K.
\section{Discussion}
The Si membrane technology is a promising approach
to realize a variety of SIS mixers at frequencies 
up to, or exceeding, 1~THz. NbTiN or other high 
gap superconducting materials will be necessary 
to provide almost lossless transport of the 
signals in these mixers. However, combining Nb SIS 
junctions with NbTiN transmission lines will need
to be done with care. Quasiparticle 
trapping will result in heating of the Nb electrodes 
of the SIS junctions. The result is a reduction of 
actual gap frequency of the material \cite{leone2001}
and a reduced performance of the mixer.

Moreover, our work suggests a possibility to produce 
on-chip balanced HEB mixers. Some of the authors of 
this paper recently showed that single-ended 
HEB devices can be reliably fabricated on 
$2~\mu$m thin $\mathrm{Si_3N_4}$ SOI wafers and that beamlead technology
can be used to hold the devices suspended inside the mixer 
block even at working frequencies as high as 
2.5~THz \cite{puetz2011}. In a recent paper from the GARD group, 
the SOI technology was proposed
in order to realize a future balanced waveguide
HEB mixer for the RF band $1.6-2.0$~THz \cite{dochev2011}.
At frequencies as high as e.g.~2~THz the question remains 
whether a waveguide branch-line coupler or a planar coupler
is the more favorable technology to realize a balanced HEB 
mixer. In a waveguide branch-line coupler the mixer chips 
have to be separately mounted which is a critical technological
step. However, it offers the opportunity to select matching pairs of
devices. A planar coupler in normal metal technology, because of
the high frequency, is fabricated with integrated mixer devices, but
offers no possibility to counteract fabrication tolerances after
device processing.
\section{Conclusion}
To conclude, we have measured the uncorrected DSB receiver noise
temperature of our device between 440~GHz and 495~GHz and obtain
values ranging from 70~K to 130~K representing state-of-the-art 
performance. An inverse polarity bias for the two mixers
does not change the performance of the device compared
to an equal bias polarity for the two mixers, provided the 
sum and the difference port of the IF hybrid is used, respectively.
The noise rejection of the device is between 6~dB and 18~dB.
Noise added to the mixer's LO port can be directly measured in
the IF output power and does not significantly increase $T_{rec,DSB}$.
The noise rejection mechanism is directly observable in
the IF output power as function of the bias voltage 
over the whole voltage range of the photon step.
\section*{Acknowledgments}
This work is funded by BMBF, Verbundforschung
Astronomie under contract number 05A08PK2 and by 
the Deutsche Forschungsgemeinschaft,
Sonderforschungsbereich
956. The excellent support of the in-house machine shop
is greatfully acknowledged.
M.P.~Westig thanks the
Bonn-Cologne Graduate School of Physics and Astronomy 
for financial support.

\end{document}